\documentclass[11pt]{article}

\usepackage[margin=1in]{geometry}
\usepackage[numbers,sort&compress]{natbib}
\usepackage{graphicx}
\usepackage{amsmath,amssymb}
\usepackage{booktabs}
\usepackage{microtype}
\usepackage{hyperref}
\usepackage{algorithm}
\usepackage{algpseudocode}
\usepackage{authblk}

\hypersetup{
  colorlinks=true,
  linkcolor=blue,
  citecolor=blue,
  urlcolor=blue
}

\title{\textbf{Constrained Enumeration Reveals Hidden Optima and Precision-Dependent Degeneracy\\ in Modularity-Based Community Detection}}

\author[1]{Fabio Morea} 
\affil[1]{Area Science Park, Trieste, Italy}
\date{March 2026}

\begin{document}
\maketitle

\begin{abstract}
Modularity landscapes are often flat near the top: many distinct partitions achieve indistinguishable scores, and heuristic restarts can still miss accessible optima. We introduce a two-phase workflow that (i) samples partitions until novelty saturates, then (ii) localises instability to a small \emph{fuzzy} subset of nodes and enumerates only that residual ambiguity under a locked stable core. Across a simple clique-ring (methodology showcase), a noisy 50-clique benchmark, and a real weighted collaboration network, constrained enumeration systematically expands plateau coverage and can improve modularity beyond extensive Louvain restarts. Finally, we show that edge-weight rounding qualitatively reshapes plateau structure, making precision sensitivity an essential part of solution-space reporting.
\end{abstract}

\section{Introduction}
Community detection is a central task in network science. Among many approaches, modularity maximisation remains popular because it is simple, scalable, and interpretable \citep{NewmanGirvan2004,Newman2006}. However, the modularity landscape is often rugged and \emph{highly degenerate}: many structurally distinct partitions can attain indistinguishable modularity values \citep{Good2010,MassenDoye2005}. Degeneracy makes interpretation difficult, because any single ``best'' partition may be an arbitrary representative of a larger plateau of near-equivalent explanations. Moreover, modularity is affected by a resolution limit, in which small communities may be merged even when they are well-defined \citep{FortunatoBarthelemy2007}.

Since exact modularity optimisation is NP-hard \citep{Brandes2008}, practical analysis relies on heuristics such as Louvain \citep{Blondel2008} and Leiden \citep{Traag2019}, often with many random restarts. Yet repeated restarts are an \emph{unprincipled} response to degeneracy. They do not answer two questions that matter for scientific inference:
\begin{enumerate}
    \item \textbf{Maximality:} how many runs are sufficient to ensure that the maximum modularity value $Q_{\max}$ has been reached?
    \item \textbf{Cardinality:} is there a single maximum or a plateau of equivalent partitions? If so, how many \emph{distinct} partitions constitute the $Q_{\max}$ plateau?
\end{enumerate}
Even on moderate graphs, the space of partitions is discrete and combinatorially enormous. There is no canonical geometry of partitions, and different similarity measures induce different, non-Euclidean structures \citep{Meila2007}. Enumeration over the full graph is therefore infeasible; heuristics are necessary. The question is how to \emph{combine} heuristics with more exhaustive tools in a way that is both computationally viable and epistemically informative.

In previous work we proposed a Bayesian framework to explore the solution space of community detection algorithms and a taxonomy of solution-space regimes \citep{MoreaDeStefano2025SciRep}. Here we extend this direction with a complementary idea: after heuristic exploration has stabilised, use the variability of solutions to localise ambiguity, lock stable structure, and then apply \emph{constrained enumeration} only to the residual uncertain subgraph. This yields a practical route to answer both questions above: maximality is addressed by focusing enumeration on the only degrees of freedom that remain uncertain, and cardinality is answered by counting distinct maximal partitions under exact partition identity.

\paragraph{Contributions.}
We make three contributions.
(i) We show empirically (synthetic and real) that constrained enumeration can reveal \emph{hidden optima}---partitions with $Q$ higher than those found by extensive heuristic restarts.
(ii) We quantify plateau multiplicity by enumerating distinct maximal partitions beyond heuristic coverage, and we interpret this as a stability diagnostic (core vs \emph{fuzzy} uncertainty).
(iii) We demonstrate that weight precision (e.g., rounding to two versus three decimals) can qualitatively reshape plateau geometry, implying that plateau analysis is essential to understand how seemingly small modelling choices impact community detection conclusions.

\section{Methods}

\subsection{Modularity and exact partition identity}
Let $G=(V,E)$ be an undirected weighted graph with weights $w_{ij}\ge 0$, total weight $m=\sum_{(i,j)\in E} w_{ij}$, and node strengths $k_i=\sum_j w_{ij}$.
For a partition $\mathcal{P}$ with community labels $c_i$, weighted modularity is
\begin{equation}
Q(\mathcal{P}) = \frac{1}{2m}\sum_{i,j}\left(w_{ij} - \frac{k_i k_j}{2m}\right)\mathbf{1}[c_i=c_j].
\label{eq:Q}
\end{equation}
We compute $Q$ using the equivalent community-sum form
\begin{equation}
Q(\mathcal{P}) = \sum_{c}\left(\frac{w_c^{\mathrm{in}}}{m} - \left(\frac{k_c}{2m}\right)^2\right),
\label{eq:Q-community}
\end{equation}
where $w_c^{\mathrm{in}}$ is the total internal weight of community $c$ (each undirected edge counted once) and $k_c=\sum_{i\in c}k_i$ is the community strength.

To treat ``new'' partitions rigorously we use a canonical relabelling via restricted growth strings (RGS). Two partitions are identical if and only if their RGS encodings match \emph{exactly} (zero tolerance, up to label permutations). This strict identity is essential for plateau cardinality.

\subsection{Phase I: heuristic exploration with Bayesian saturation}
Phase I repeatedly runs a stochastic heuristic $H$ (Louvain in our experiments) under random vertex permutations, producing a sequence of partitions $\{\pi_t\}$. Let
\begin{equation}
X_t=\mathbf{1}[\mathrm{RGS}(\pi_t)\ \text{is unseen in}\ \{\pi_1,\dots,\pi_{t-1}\}]
\end{equation}
be the indicator that run $t$ produces a \emph{new} partition. We model novelty as Bernoulli with rate $p_{\mathrm{new}}$, with a Beta prior $p_{\mathrm{new}}\sim\mathrm{Beta}(1,1)$. After $t$ runs the posterior is $\mathrm{Beta}(1+n_{\mathrm{new}},\,1+n_{\mathrm{old}})$ where $n_{\mathrm{new}}=\sum_{s\le t}X_s$.

We stop heuristic exploration when the posterior indicates that further runs are unlikely to discover new partitions \emph{and} uncertainty is small:
\begin{equation}
\Pr(p_{\mathrm{new}}<\tau)\ge \alpha
\quad\text{and}\quad
\big(q^{u}_{\alpha}-q^{\ell}_{\alpha}\big)\le \omega,
\label{eq:stop_rule}
\end{equation}
where $q^{\ell}_{\alpha}$ and $q^{u}_{\alpha}$ are the lower/upper $\alpha$-credible quantiles of the posterior, $\tau$ is a novelty tolerance, and $\omega$ controls the required tightness of the credible interval.
This provides a defensible stopping rule that replaces ad hoc run counts.

Phase I can also be used as a diagnostic of the \emph{solution-space regime}, using the taxonomy proposed in \citep{MoreaDeStefano2025SciRep} (e.g., dominant vs multiple solutions). In this paper we focus on cases where the Phase I ensemble suggests plateau-like multiplicity and motivates Phase II completion.

\subsection{Phase II: core--fuzzy decomposition and constrained enumeration}
From the Phase I ensemble we compute a co-occurrence matrix $C$, where $C_{ij}$ is the fraction of sampled partitions in which nodes $i$ and $j$ co-cluster. Using a stability threshold $s$, we separate nodes into a stable \emph{core} (consistently co-clustered structure) and an ambiguous \emph{fuzzy} set (nodes whose membership varies across high-$Q$ solutions). The fuzzy-induced subgraph is then decomposed into connected components $\{\mathcal{F}_1,\dots,\mathcal{F}_r\}$. Component decomposition is crucial: disconnected fuzzy components can be enumerated independently and combined by Cartesian product.

\paragraph{Locked core without label dependence.}
Enumeration requires a fixed reference \emph{core structure} to define which community each core node belongs to. Because we use RGS canonical labelling, community \emph{IDs} are not meaningful: they are arbitrary up to permutation and are normalised away by the RGS map. The role of the reference partition is therefore not ``to choose the best labels'', but to provide a concrete representative of the \emph{core block structure} induced by the ensemble.

In practice we select a representative partition $\pi^{\star}$ from the top plateau (or near-top set) and lock the induced core communities. Any two representatives that agree on the core induce the same locked constraints (up to relabelling), and hence the same enumerated set after RGS deduplication. When multiple candidates differ slightly on the core, we choose $\pi^{\star}$ by maximising core stability (equivalently, consistency with the co-occurrence matrix); ties may be broken by modularity.

For each fuzzy component $\mathcal{F}_i$, we enumerate feasible assignments while keeping the core constraints fixed. In this paper we report only an \texttt{improve\_only} mode: we keep candidates with
\begin{equation}
Q > Q_{\mathrm{bench}},
\end{equation}
where $Q_{\mathrm{bench}}$ is a pragmatic Phase I baseline such as the mean modularity over Phase I runs, $Q_{\mathrm{bench}}=\mathbb{E}[Q]$, or (when desired) the best heuristic value observed in Phase I.
To keep enumeration practical, we compute modularity via incremental updates of $(k_c,w_c^{\mathrm{in}})$ using precomputed graph arrays, avoiding full recomputation for each candidate.

\begin{algorithm}[t]
\caption{Two-phase plateau characterisation (CE-CD)}
\label{alg:cecd}
\begin{algorithmic}[1]
\State \textbf{Input:} graph $G$, heuristic $H$, thresholds $(\tau,\alpha,\omega)$, stability threshold $s$, baseline $Q_{\mathrm{bench}}$.
\State \textbf{Phase I (Bayesian saturation):} run $H$ repeatedly; record partitions; deduplicate by RGS; stop when Eq.~(\ref{eq:stop_rule}) holds.
\State \textbf{Phase II (core--fuzzy + enumeration):} compute co-occurrence matrix $C$ over the Phase I ensemble; identify stable \emph{core} and ambiguous \emph{fuzzy} using $s$.
\State Decompose the fuzzy-induced subgraph into connected components $\{\mathcal{F}_1,\dots,\mathcal{F}_r\}$.
\State Choose a representative $\pi^{\star}$ that maximises core stability (tie-break by $Q$) and lock the induced core constraints.
\State For each component $\mathcal{F}_i$, enumerate feasible assignments under locked-core constraints; evaluate $Q$ via incremental updates; keep only candidates with $Q>Q_{\mathrm{bench}}$.
\State Combine component-wise assignments; deduplicate by exact RGS; return the final solution set (\texttt{improve\_only}).
\end{algorithmic}
\end{algorithm}

\section{Experiments}
We report four experiments. Experiment~1 is intentionally easy and serves to \emph{showcase the methodology}: Phase~I rapidly stabilises, the core--fuzzy split is visually interpretable, and plateau structure can be read directly. Experiment~2 scales the same construction to a noisy 50-clique benchmark where degeneracy is substantial and heuristic restarts under-sample the top plateau. Experiment~3 applies the workflow to a real weighted collaboration network, highlighting how the core--fuzzy decomposition yields a stable summary even when multiple near-optimal solutions exist. Experiment~4 demonstrates that plateau geometry is \emph{precision-dependent}: rounding weights can merge or split plateaus and change what appears ``optimal''.

\subsection{Experiment 1: clique-ring toy example (methodology showcase)}
We begin with a small ring of five cliques (six nodes each) with a handful of weakly attached or bridging nodes. The goal is not difficulty, but \emph{interpretability}: the mesoscale structure is obvious, so any residual ambiguity can be attributed to near-degeneracy rather than to a genuinely unclear community signal.

\begin{figure}[h]
    \centering
    \includegraphics[width=1\linewidth]{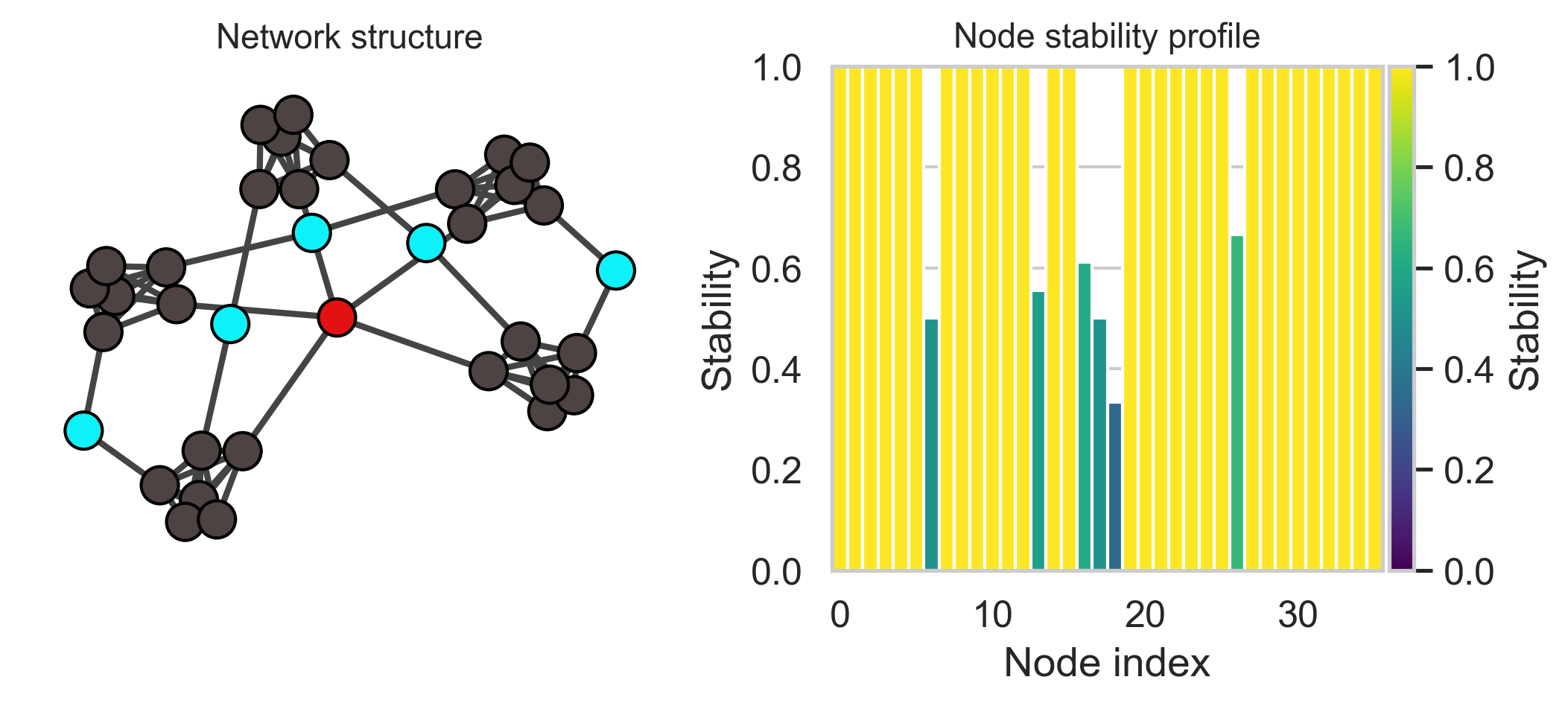}
    \caption{\textbf{Experiment 1 (toy): core--fuzzy localisation via node stability.} \emph{Left:} clique-ring toy network with weakly attached/boundary nodes. \emph{Right:} node stability (co-clustering consistency across Phase~I high-$Q$ solutions). Most nodes form a stable core (near stability 1), while ambiguity concentrates on a small fuzzy subset, motivating constrained enumeration on only those degrees of freedom.}
    \label{fig:toy-stability}
\end{figure}

Phase~I (stochastic Louvain restarts with Bayesian novelty saturation) stabilises quickly in this toy setting. We then compute node stability and identify a large stable core together with a small \emph{fuzzy} subset whose assignments vary across high-$Q$ solutions. Figure~\ref{fig:toy-stability} visualises this localisation: instability concentrates on a few boundary nodes rather than spreading across the graph.

\begin{figure}[h]
    \centering
    \includegraphics[width=1\linewidth]{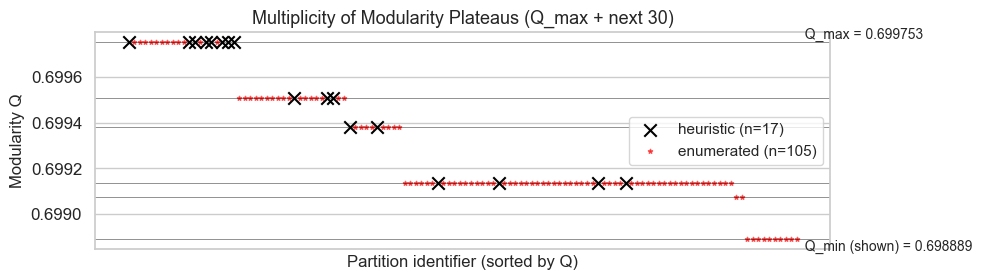}
    \caption{\textbf{Experiment 1 (toy): plateau multiplicity---heuristics vs constrained enumeration.} Distinct partitions are ordered by decreasing modularity $Q$ (horizontal axis). Black crosses are unique partitions discovered by Phase~I heuristic sampling; red stars are additional partitions found by constrained enumeration restricted to the fuzzy component(s) under a locked core. Even when heuristics reach $Q_{\max}$, enumeration reveals additional distinct partitions on the top plateau, turning an apparent single answer into an explicit multiplicity.}
    \label{fig:toy-plateaus}
\end{figure}

Finally, Figure~\ref{fig:toy-plateaus} compares plateau sampling from heuristics against constrained enumeration on the fuzzy component(s) under a locked core. Even here---where heuristics typically reach $Q_{\max}$---enumeration reveals that heuristic restarts provide only a partial view of plateau multiplicity, clarifying how many distinct explanations exist at essentially the same score.

\subsection{Experiment 2: noisy 50-clique benchmark (hidden optima beyond heuristic baselines)}
We next analyse a substantially harder synthetic graph: a ring of 50 cliques augmented with (i) bridge nodes connecting successive cliques, (ii) a central connector node linked to the ring, and (iii) sparse random inter-clique noise edges. The ``ground truth'' clique structure is clear at a coarse level, but modularity becomes \emph{degenerate} because many boundary and bridge assignments have negligible effect on $Q$ \citep{FortunatoBarthelemy2007,Good2010}.

We run Phase~I until novelty saturates, and we use the resulting ensemble to localise a stable core and a small fuzzy region. We then perform constrained enumeration on the fuzzy components, reporting only candidates that exceed a Phase~I baseline $Q_{\mathrm{bench}}$ (here we take $Q_{\mathrm{bench}}=\mathbb{E}[Q]$ over Phase~I runs, which acts as a pragmatic benchmark for what repeated heuristic exploration typically achieves).

\begin{figure}[h]
    \centering
    \includegraphics[width=.8\linewidth]{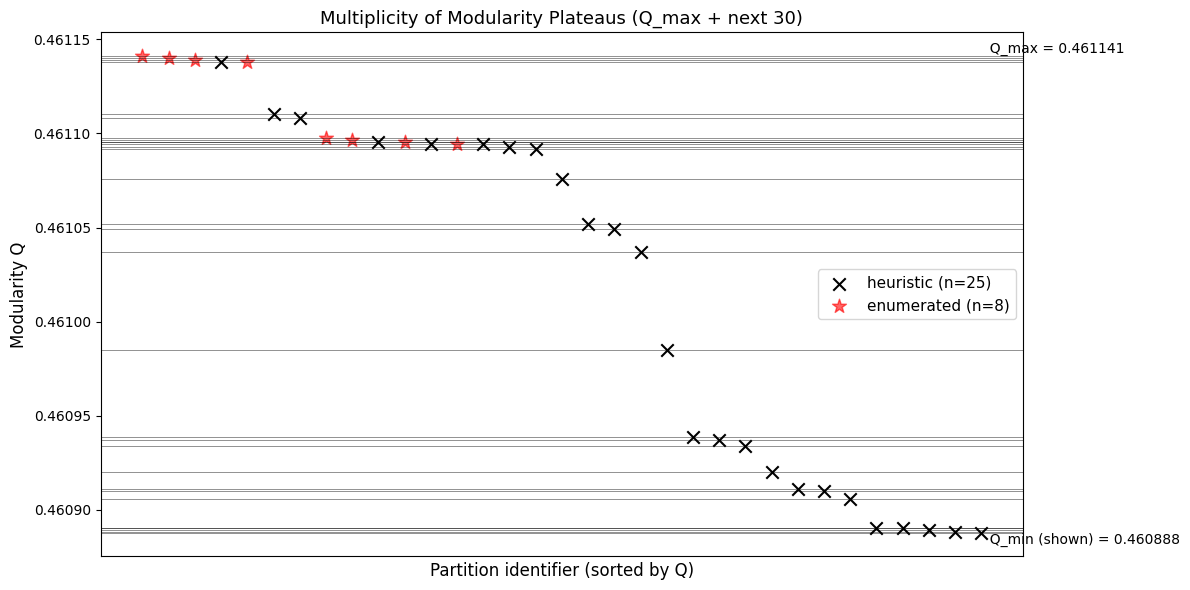}
    \caption{\textbf{Experiment 2 (50 cliques): hidden optima and expanded plateau coverage.} Each point is a distinct partition (exact RGS identity), ordered by decreasing $Q$. Black crosses are unique partitions obtained by heuristic restarts (Phase~I); red stars are partitions found by constrained enumeration on fuzzy components (Phase~II) that exceed the Phase~I baseline $Q_{\mathrm{bench}}$ (here, $\mathbb{E}[Q]$ over Phase~I runs). Enumeration both densifies the top plateau (revealing under-sampled multiplicity) and can produce partitions that outperform restart-based exploration.}
    \label{fig:bench-plateaus}
\end{figure}

Figure~\ref{fig:bench-plateaus} shows two effects. First, heuristic restarts under-sample the top plateau: many distinct high-$Q$ partitions exist but are rarely returned. Second, constrained enumeration can yield \emph{hidden optima}, i.e., partitions that improve upon the heuristic baseline and may even exceed the best value observed after extensive restarts. The point is not the absolute gain (which can be small in plateau regimes), but the methodological consequence: once ambiguity is localised, targeted enumeration can systematically close the gap left by restart-based exploration.

\clearpage
\subsection{Experiment 3: real weighted collaboration network (Horizon projects)}
We analyse a real weighted collaboration network derived from Horizon programme data, introduced in \citep{MoreaSoraciDeStefano2025ORE}. This network is weighted and heterogeneous, and community structure is expected to be \emph{fuzzy}: different granularities and boundary assignments can plausibly coexist. This makes it an appropriate test for stability and plateau interpretation in practice.

\begin{figure}[h]
    \centering
    \includegraphics[width=.5\linewidth]{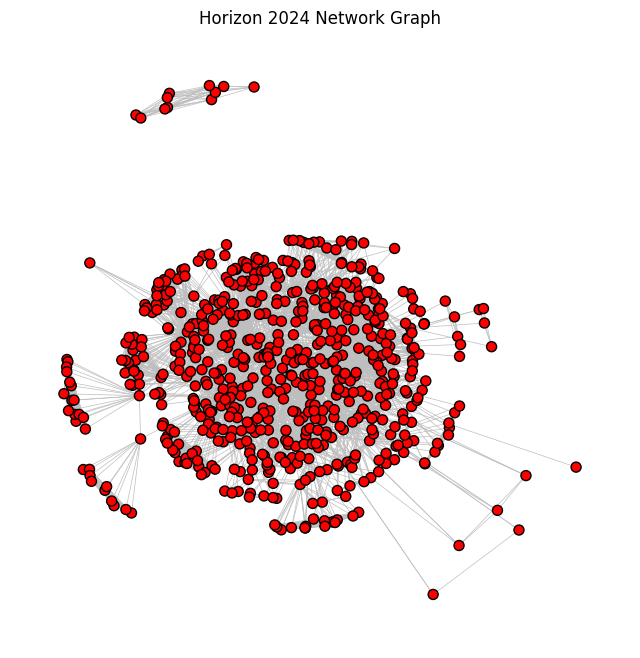}
    \caption{\textbf{Experiment 3 (real network): weighted Horizon collaboration graph.} Weighted edges encode collaboration strength (data from \citep{MoreaSoraciDeStefano2025ORE}). In heterogeneous weighted networks, many near-optimal partitions can coexist; this motivates reporting solution-space structure (plateaus and stability) rather than a single partition.}
    \label{fig:real-graph}
\end{figure}

\begin{figure}[h]
    \centering
    \includegraphics[width=.8\linewidth]{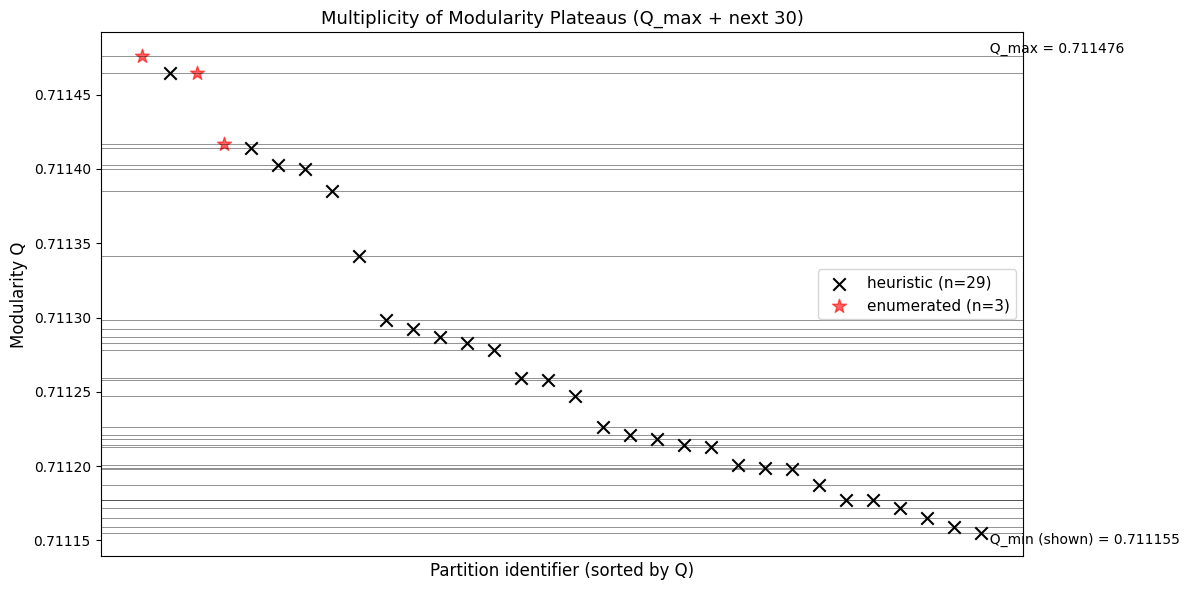}
    \caption{\textbf{Experiment 3 (real network): constrained enumeration improves and stabilises the top solution set.} Black crosses are distinct partitions from Phase~I heuristic exploration; red stars are additional partitions from Phase~II enumeration over fuzzy components under a locked core, filtered to those exceeding the Phase~I baseline $Q_{\mathrm{bench}}$. Beyond improving $Q$, the workflow isolates residual ambiguity to a small fuzzy subset, yielding a stable core partition for interpretation.}
    \label{fig:real-plateaus}
\end{figure}

Figure~\ref{fig:real-plateaus} shows that the same pattern seen in Experiment~2 persists in real data: heuristic exploration produces a small set of distinct top solutions, while constrained enumeration adds additional partitions at the top plateau and can reach a higher modularity value than repeated Louvain alone. Importantly, the output is not merely ``more partitions'': the core--fuzzy decomposition exposes which parts of the partition are stable across the near-optimal set, yielding a stable summary even when the top plateau contains multiple configurations.

\begin{figure}[h]
    \centering
    \includegraphics[width=\linewidth]{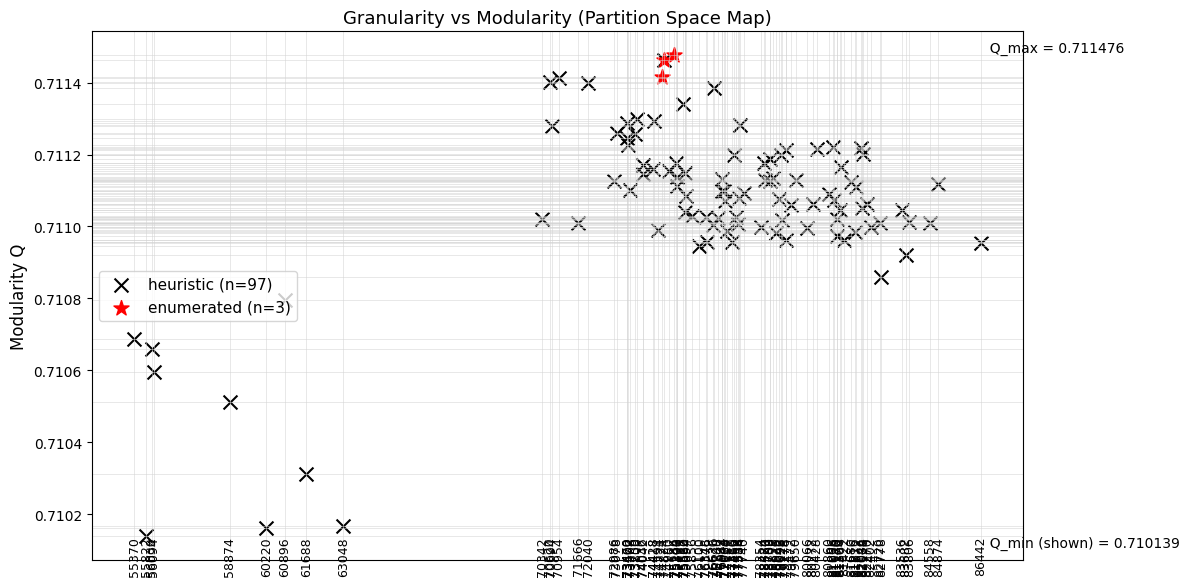}
    \caption{\textbf{Experiment 3 (real network): solution-space map (modularity $Q$ vs number of communities $\Gamma$).} Each point is a distinct partition (exact RGS identity). Horizontal bands indicate degeneracy plateaus in $Q$, while variation in $\Gamma$ reveals competing granularities among near-optimal solutions. This map distinguishes ``boundary ambiguity at fixed granularity'' from ``alternative resolutions with comparable support'', complementing the solution-space perspective in \citep{MoreaDeStefano2025SciRep}.}
    \label{fig:real-psm}
\end{figure}

To connect plateau multiplicity with granularity, Figure~\ref{fig:real-psm} plots modularity against $\Gamma$ (number of communities) for the same solution set. The resulting ``solution-space map'' highlights an important point: degeneracy is not only about ties at a single value of $Q$, but also about competition between \emph{near-optimal granularities}. In practice, a method that reports only one partition obscures whether a competing granularity is nearly as plausible. Plateau analysis makes these alternatives explicit and therefore supports more defensible interpretation.

\clearpage
\subsection{Experiment 4: weight precision reshapes plateau geometry}
Finally, we repeat the analysis of Experiment~3 under different rounding schemes for edge weights (two-decimal vs three-decimal rounding). This isolates the influence of weight precision on plateau geometry and on the apparent uniqueness of the optimal solution.

\begin{figure}[h]
    \centering
    \includegraphics[width=.8\linewidth]{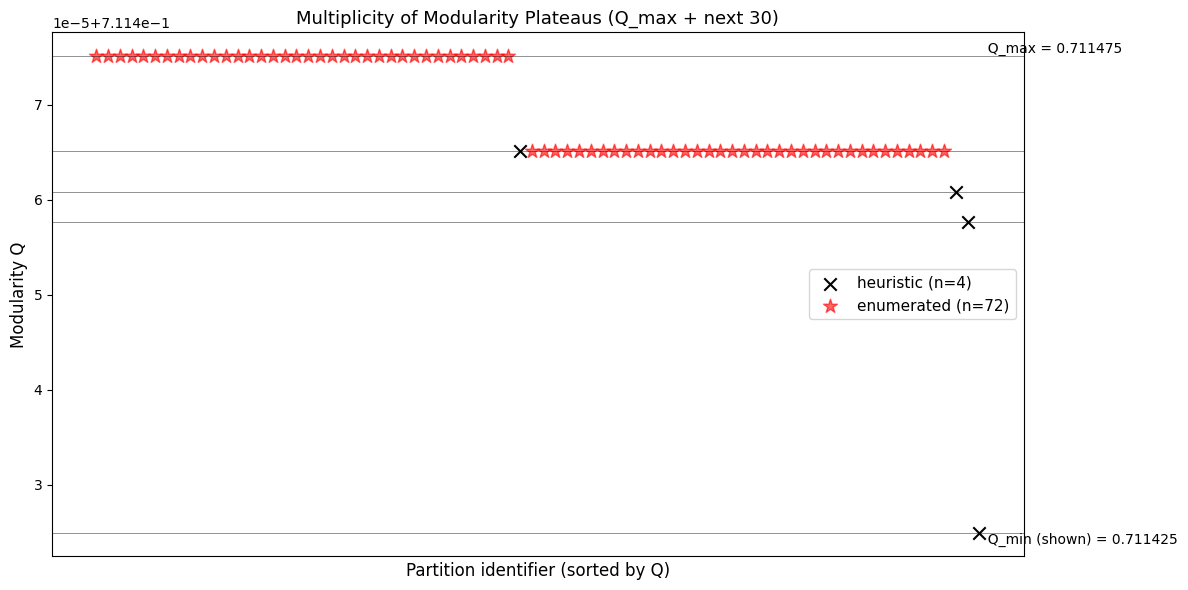}
    \caption{\textbf{Experiment 4 (precision sensitivity): weight rounding reshapes plateau geometry.} Plateau multiplicity after rounding edge weights (e.g., to two decimals) on the real network. Coarser precision merges many near-ties into wide flat plateaus, increasing the number of distinct maximal (or near-maximal) configurations; finer precision can split these plateaus into a stratified ordering. This demonstrates that degeneracy and ``optimality'' can be artefacts of measurement precision, and should be reported as part of solution-space characterisation.}
    \label{fig:rounding}
\end{figure}

Figure~\ref{fig:rounding} shows that rounding weights can qualitatively reshape the top of the modularity landscape. With two-decimal rounding, the top solutions can collapse into wide flat plateaus, with many distinct partitions sharing identical modularity. With higher-precision weights, some ties are broken into a fine stratification of $Q$ values. This observation has a direct epistemic implication: if the data do not support fine weight precision, then an apparent unique optimum at high precision may be an artefact of micro-perturbations rather than a meaningful structural distinction. Plateau analysis is therefore necessary to understand how modelling choices (such as rounding) affect conclusions about the ``best'' community structure.

\clearpage

\section{Discussion}
The experiments support the conclusion that in degenerate modularity landscapes, it is insufficient to report a single partition and a single modularity value. The scientifically relevant characteristics to report are (i) the maximum modularity value reached relative to a defensible baseline, and (ii) the set (and size) of partitions that attain the top plateau (or exceed the baseline).

Full enumeration is infeasible on the full graph, but heuristics can be used to \emph{localise} the search: stable structure is locked, and the remaining ambiguity is enumerated component-wise. This yields a principled bridge between stochastic exploration and local completeness.

The weight-rounding experiment adds an additional layer: degeneracy is partly \emph{precision-dependent}. When modularity differences are on the order of $10^{-5}$, the difference between two- and three-decimal weights can decide whether plateaus appear as true ties or as an arbitrary strict ordering. This suggests that community detection pipelines should treat weight precision as a modelling assumption and should report the robustness of plateaus under plausible precision regimes.

Although our implementation uses Louvain in Phase I, the argument is method-agnostic. Any stochastic community detection procedure that produces an ensemble of partitions can be paired with core--fuzzy localisation and constrained enumeration. 

\section{Conclusion}
We showed that constrained enumeration is a practical and informative complement to heuristic modularity optimisation. It can reveal higher optima than extensive restarts, quantify plateau multiplicity, and provide stable summaries through core--fuzzy separation. We further showed that plateau geometry depends on weight precision, making plateau analysis essential for understanding the effect of modelling choices on community detection results.

\section*{Data and code availability}
Code and reproducible notebooks will be made available upon publication.

\bibliographystyle{unsrtnat}
\bibliography{references}

@article{NewmanGirvan2004,
  title        = {Finding and evaluating community structure in networks},
  author       = {Newman, Mark E. J. and Girvan, Michelle},
  journal      = {Physical Review E},
  volume       = {69},
  number       = {2},
  pages        = {026113},
  year         = {2004},
  doi          = {10.1103/PhysRevE.69.026113}
}

@article{Newman2006,
  title        = {Modularity and community structure in networks},
  author       = {Newman, Mark E. J.},
  journal      = {Proceedings of the National Academy of Sciences},
  volume       = {103},
  number       = {23},
  pages        = {8577--8582},
  year         = {2006},
  doi          = {10.1073/pnas.0601602103}
}

@article{FortunatoBarthelemy2007,
  title        = {Resolution limit in community detection},
  author       = {Fortunato, Santo and Barth{\'e}lemy, Marc},
  journal      = {Proceedings of the National Academy of Sciences},
  volume       = {104},
  number       = {1},
  pages        = {36--41},
  year         = {2007},
  doi          = {10.1073/pnas.0605965104}
}

@article{MassenDoye2005,
  title        = {Identifying communities within networks: the energy landscape of modularity},
  author       = {Massen, C. P. and Doye, J. P. K.},
  journal      = {Physical Review E},
  volume       = {71},
  number       = {4},
  pages        = {046101},
  year         = {2005},
  doi          = {10.1103/PhysRevE.71.046101}
}

@article{Good2010,
  title        = {Performance of modularity maximization in practical contexts},
  author       = {Good, Benjamin H. and de Montjoye, Yves-Alexandre and Clauset, Aaron},
  journal      = {Physical Review E},
  volume       = {81},
  number       = {4},
  pages        = {046106},
  year         = {2010},
  doi          = {10.1103/PhysRevE.81.046106}
}

@article{Blondel2008,
  title        = {Fast unfolding of communities in large networks},
  author       = {Blondel, Vincent D. and Guillaume, Jean-Loup and Lambiotte, Renaud and Lefebvre, Etienne},
  journal      = {Journal of Statistical Mechanics: Theory and Experiment},
  volume       = {2008},
  number       = {10},
  pages        = {P10008},
  year         = {2008},
  doi          = {10.1088/1742-5468/2008/10/P10008}
}

@article{Traag2019,
  title        = {From {L}ouvain to {L}eiden: guaranteeing well-connected communities},
  author       = {Traag, Vincent A. and Waltman, Ludo and van Eck, Nees Jan},
  journal      = {Scientific Reports},
  volume       = {9},
  number       = {1},
  pages        = {5233},
  year         = {2019},
  doi          = {10.1038/s41598-019-41695-z}
}

@article{Brandes2008,
  title        = {On modularity clustering},
  author       = {Brandes, Ulrik and Delling, Daniel and Gaertler, Marco and G{\"o}rke, Robert and Hoefer, Martin and Nikoloski, Zoran and Wagner, Dorothea},
  journal      = {IEEE Transactions on Knowledge and Data Engineering},
  volume       = {20},
  number       = {2},
  pages        = {172--188},
  year         = {2008},
  doi          = {10.1109/TKDE.2007.190689}
}

@article{Meila2007,
  title        = {Comparing clusterings---an information based distance},
  author       = {Meil{\u a}, Marina},
  journal      = {Journal of Multivariate Analysis},
  volume       = {98},
  number       = {5},
  pages        = {873--895},
  year         = {2007},
  doi          = {10.1016/j.jmva.2006.11.013}
}

@article{MoreaSoraciDeStefano2025ORE,
  author       = {Morea, Fabio and Soraci, Alberto and De Stefano, Domenico},
  title        = {Mapping leadership and communities in {EU}-funded research through network analysis},
  journal      = {Open Research Europe},
  year         = {2025},
  volume       = {4},
  pages        = {268},
  doi          = {10.12688/openreseurope.18544.2},
  url          = {https://open-research-europe.ec.europa.eu/articles/4-268/v2}
}

@article{MoreaDeStefano2025SciRep,
  author       = {Morea, Fabio and De Stefano, Domenico},
  title        = {A comprehensive framework for solution space exploration in community detection},
  journal      = {Scientific Reports},
  year         = {2025},
  volume       = {15},
  pages        = {38148},
  doi          = {10.1038/s41598-025-22046-7},
  url          = {https://www.nature.com/articles/s41598-025-22046-7}
}

\end{document}